\UseRawInputEncoding
\documentclass[%
superscriptaddress,
 amsmath,amssymb,
 aps,
 pra,
]{revtex4-2}

\usepackage{graphicx}
\usepackage{dcolumn}
\usepackage{bm}
\usepackage[table]{xcolor}
\definecolor{green}{HTML}{9a94bc}
\definecolor{myColor}{HTML}{9A94BC}

\usepackage{color}
\usepackage{booktabs}

\begin{document}

\preprint{APS/123-QED}

\title{Towards a monolithic platform for coupling superconducting circuits to low-loss microwave phonons in AlScN on 4H-SiC}

\author{Yuanchen Deng}
    \thanks{These authors contributed equally to this work}
    \affiliation{Department of Electrical, Computer and Energy Engineering, University of Colorado Boulder, Boulder, Colorado 80309, USA}%
\author{William W. Roberts}%
    \thanks{These authors contributed equally to this work}
    \affiliation{Department of Electrical, Computer and Energy Engineering, University of Colorado Boulder, Boulder, Colorado 80309, USA}%
\author{Sueli Skinner-Ramos}
     \thanks{These authors contributed equally to this work}
    \affiliation{Microsystems Engineering, Science, and Applications, Sandia National Laboratories, Albuquerque, NM, USA}%
\author{Dalton Anderson}%
    \affiliation{College of Optical Sciences, University of Arizona, Tucson, AZ, USA}%
\author{Katherine Hewey}%
    \affiliation{College of Optical Sciences, University of Arizona, Tucson, AZ, USA}%
\author{Xingyu Du}
    \affiliation{Department of Electrical and Systems Engineering, University of Pennsylvania, Philadelphia, Pennsylvania 19104, USA}%
\author{Michael Miller}
    \affiliation{Microsystems Engineering, Science, and Applications, Sandia National Laboratories, Albuquerque, NM, USA}%
\author{Brandon Smith}
    \affiliation{Microsystems Engineering, Science, and Applications, Sandia National Laboratories, Albuquerque, NM, USA}%
\author{Hwijong Lee}
    \affiliation{Center for Integrated Nanotechnologies, Sandia National Laboratories, Albuquerque, NM, USA}%
\author{Pingping Chen}
    \affiliation{Department of Electrical, Computer and Energy Engineering, University of Colorado Boulder, Boulder, Colorado 80309, USA}%
\author{Charles Thomas Harris}
     \affiliation{Center for Integrated Nanotechnologies, Sandia National Laboratories, Albuquerque, NM, USA}%
\author{Roy H. Olsson III}
    \affiliation{Department of Electrical and Systems Engineering, University of Pennsylvania, Philadelphia, Pennsylvania 19104, USA}%
\author{Lisa Hackett}
    \affiliation{Microsystems Engineering, Science, and Applications, Sandia National Laboratories, Albuquerque, NM, USA}%
\author{Rupert Lewis}
    \affiliation{Division of Material, Physical, and Chemical Sciences, Sandia National Laboratories, Albuquerque, NM, USA}%
\author{Matt Eichenfield}
\email{matt.eichenfield@colorado.edu}
    \affiliation{Department of Electrical, Computer and Energy Engineering, University of Colorado Boulder, Boulder, Colorado 80309, USA}%
    \affiliation{Center for Integrated Nanotechnologies, Sandia National Laboratories, Albuquerque, NM, USA}%
    \affiliation{Microsystems Engineering, Science, and Applications, Sandia National Laboratories, Albuquerque, NM, USA}%
    \affiliation{College of Optical Sciences, University of Arizona, Tucson, AZ, USA}%

\date{\today}

\begin{abstract}
Hybrid superconducting-phonon quantum information processing is promising for cavity QED, measurement-based quantum computing, and many other quantum applications. This is because phonons, relative to microwave photons at the same frequency, can in principle provide ultra-compact footprints, extremely low losses, and higher degrees of connectivity. Moreover, phonons can be strongly coupled to superconducting circuits through the piezoelectric effect. However, this promise rests on finding \textit{scalable} platforms that can achieve all these benefits without degrading superconducting circuit performance. This strongly motivates a material platform that \textit{monolithically} combines low phononic loss, strong electromechanical coupling, and superconducting-circuit compatibility, without requiring suspended phononics for low loss. Here, we propose and characterize a monolithic quantum acoustic platform that combines aluminum superconducting circuits on exposed silicon carbide (SiC) with piezoelectric aluminum scandium nitride (AlScN) on SiC for integrated phononics. This architecture is enabled by the selective removal of AlScN from designated regions of the chip, allowing aluminum superconducting microwave resonators to be fabricated directly on the SiC while preserving adjacent AlScN-on-SiC regions for phonon transduction. The resulting Al-on-SiC resonators exhibit a coherent lifetime of 2.9 $\mu$s, demonstrating the compatibility of exposed SiC regions with aluminum superconducting quantum devices. In parallel, cryogenic surface acoustic delay-line measurements on the retained AlScN-on-SiC regions possess low phononic propagation loss at 4.05 GHz, corresponding to an estimated phonon lifetime of 7.6 $\mu$s. Together with the previously demonstrated electromechanical coupling coefficient of approximately 4.3 $\%$ and a theoretical upper bound of 8$\%$ in this frequency range, these results establish Al-on-SiC/AlScN-on-SiC as a promising monolithic platform for integrating superconducting microwave circuits with piezoelectric phononic components for quantum acoustic networking and hybrid quantum systems.

\end{abstract}

\maketitle


\section{Introduction}

In the past decade, significant development of superconducting qubits has made them a prime candidate for fault tolerant general purpose quantum computing\cite{Krantz2019AQubits}. While benefiting from fast gate times, superconducting quantum computing is still bottlenecked by its short coherence times, limited connectivity, and large device sizes\cite{Siddiqi2021EngineeringQubits}. Transition frequencies in the gigahertz range and the requirement of cryogenic temperatures inhibit coupling superconducting qubits to room temperature quantum devices such as trapped ions. One promising solution to these challenges is circuit quantum acoustodynamics (cQAD)\cite{Chu2017QuantumQubits}, where a quantum architecture is constructed with properly coupled superconducting qubits and acoustic resonators. Such hybrid architectures combine the exceptional controllability and strong intrinsic nonlinearity of superconducting qubits with the long coherence times enabled by low-loss acoustic resonators. These cQAD architectures have demonstrated phonon-mediated quantum state transfer and remote qubit entanglement\cite{Bienfait2019Phonon-mediatedEntanglement,Wollack2022QuantumResonators}, quantum erasure with phonons\cite{Bienfait2020QuantumPhonons}, multi-phonon Fock states\cite{Chu2018CreationResonator,Sletten2019ResolvingQubit}, quantum control of surface acoustic wave (SAW) phonons\cite{Satzinger2018QuantumPhonons} and resonator based quantum computing\cite{Yang2026MechanicalComputing}. In addition to cQAD architectures, phonons also offer a promising resource for measurement-based quantum computing (MBQC), where computation relies on measurements of entangled propagating resource states\cite{Qiao2023SplittingComputing,Dumur2024AApplications,Chou2025DeterministicSubstrates,Wang2025QuantumRouting,Qiao2025AcousticDetection}. Together, these architectures highlight the distinct advantages of quantum acoustics for scalable quantum information processing. Owing to their orders-of-magnitude lower propagation speed, microwave-frequency phonons have compact wavelengths that enable dense on-chip devices and long delay lines. When implemented in a low-loss material platform that supports long-lived and well-defined propagating modes, microwave-frequency phonons can support passive routing, quantum storage, and wire-free connectivity between superconducting quantum nodes.

In these quantum acoustic architectures, efficient phonon generation, transmission, and readout are usually achieved with piezoelectric materials. However, these materials often add additional loss channels that degrade the superconducting qubit coherence time. One approach to mitigate this issue is to fabricate the acoustic resonator and superconducting circuit on separate substrates and couple them via flip-chip bonding \cite{Satzinger2018QuantumPhonons,Kitzman2023PhononicQubit,Kitzman2023QuantumPhonons} or inductive coupling\cite{Qiao2023SplittingComputing,Chou2025DeterministicSubstrates,Wang2025QuantumRouting,Qiao2025AcousticDetection}. While effective, these methods add significant fabrication complexity and pose potential challenges for large-scale integration and long-term reliability. Alternatively, monolithic quantum acoustic architectures have been explored using both single material acoustic substrates and heterostructure platforms. Single material platforms, including GaAs\cite{Moores2018CavityRegime,Sletten2019ResolvingQubit} and bulk quartz\cite{Manenti2017CircuitWaves}, have enabled important demonstrations of quantum acoustic coupling. Both GaAs and quartz are intrinsically piezoelectric, but their relatively low electromechanical coupling coefficients can limit phonon generation and detection efficiency. In addition, the low dielectric constant of quartz leads to huge superconducting circuit dimensions, reducing compactness and scalability in integrated cQAD systems. Heterostructure approaches address some of these limitations by combining a piezoelectric thin film with a substrate suitable for superconducting devices. The piezoelectric layer is selectively removed to expose the low-loss substrate for superconducting circuit fabrication. Representative examples include AlN on silicon\cite{OConnell2010QuantumResonator}, AlN on sapphire\cite{Crump2023CouplingQubits}, GaN on silicon\cite{Kervinen2018InterfacingPhonons}, epitaxially grown AlGaN/GaN/NbN on SiC\cite{Gokhale2020EpitaxialAcoustodynamics} and thin-film LiNbO$_3$(TFLN) on silicon\cite{Arrangoiz-Arriola2018CouplingCavity,Arrangoiz-Arriola2019ResolvingOscillator,Wollack2022QuantumResonators,Lee2023StrongQubit,Cleland2024StudyingSensor}. Nevertheless, these heterostructure platforms also involve performance tradeoffs. AlN and GaN offer high acoustic velocities, but their relatively weak piezoelectric coefficients can limit phonon generation and detection efficiency. High-quality GaN often requires epitaxial growth or high temperature MOCVD processes, which can increase fabrication cost and impose challenges in cross-wafer uniformity and material flexibility. Epitaxially grown AlGaN/GaN/NbN heterostructures can provide high-quality interfaces and reduced defect densities, but the required epitaxial growth is costly and demands precise control over multilayer thicknesses and material properties, with cross-wafer uniformity remaining an additional challenge. TFLN has become a leading platform for strong piezoelectric coupling, but its integration generally relies on bonding or layer transfer processes rather than direct deposition or epitaxial growth, which limits the cross-wafer uniformity in heterogeneous integration. In addition, the strong in-plane anisotropy of LiNbO$_3$ causes the phase velocity, electromechanical coupling, and mode confinement to vary substantially with propagation direction, complicating the design of bends, junctions, and densely routed large-scale phononic circuits. Meanwhile, reported cryogenic phononic losses in TFLN remain relatively large for applications requiring long-lived microwave phonons. For example, Mayor \textit{et al.} reported a propagation loss of $0.7\pm0.2~\mathrm{dB/mm}$ near $3.5~\mathrm{GHz}$ at $4~\mathrm{K}$ \cite{Mayor2021GigahertzSapphire}, corresponding to a phonon lifetime on the order of $1{-}2~\mu\mathrm{s}$. More recently, Lin \textit{et al.} reported a loss of approximately $2~\mathrm{dB/mm}$ in suspended TFLN at $4~\mathrm{K}$ \cite{Lin2026ExperimentalNiobate}, yielding a sub-microsecond phonon lifetime. These values highlight the need for alternative piezoelectric material platforms that combine strong electromechanical coupling with substantially lower phononic losses at cryogenic temperatures, while providing compatibility with superconducting circuits and scalable thin-film processing.

Here we propose a new quantum acoustic material platform: high-Sc-content aluminum scandium nitride (AlScN) thin films directly grown on high resistivity silicon carbide (SiC) substrates for monolithic quantum acoustic architectures. The proposed platform provides unique performance benefits including: 1) A stronger piezoelectric coupling to surface acoustic modes compared to other deposited piezoelectric thin films via high Sc concentration AlScN. 2) A potentially lower cryogenic phononic loss enabling a microsecond-level microwave phonon lifetime. 3) A significantly enhanced phononic Kerr nonlinearity at cryogenic temperatures for potential applications like quantum phononic oscillators and parametric amplifiers.\cite{Behera2026CryogenicAlScN/SiC} 4) The SiC substrate provides a very high acoustic velocity contrast to ensure a vertical confinement of the surface phonons in AlScN. 5) The close lattice match between SiC and AlN enables direct sputter deposition of AlScN, where AlN seed layers and graded scandium-concentration interface films\cite{Du2024NearSiC} permit high Sc concentrations with large electromechanical coupling, excellent cross-wafer uniformity\cite{Deng2025MonolithicDevices}, and low defect density. This growth process is also substantially simpler and less costly than bonded platforms such as TFLN on sapphire, which is critical for scalable quantum architectures. 6) The polar in-plane isotropy of the material platform simplifies phonon-routing and phononic component design. 7) SiC offers outstanding thermal conductivity\cite{Wei2013ThermalCrystals} and provides a predominantly spin-zero host lattice for well-known spin and vacancy-center defects with long coherence times\cite{Castelletto2020SiliconApplications}. This new platform is capable of minimizing material-induced decoherence and enhancing the qubit-phonon interaction strength for advanced quantum acoustic applications. 

In this work, we demonstrate the potential of 42 $\%$ Sc-concentration AlScN grown on 4H-SiC as a monolithic superconducting quantum acoustic platform by showing that long $T_1$ time superconducting resonators and phononic waveguides can coexist on a selectively processed heterostructure. We first establish an integration route in which the AlScN layer is selectively removed to expose the underlying SiC, enabling aluminum coplanar waveguide (CPW) superconducting resonators to be fabricated directly on the exposed substrate while preserving adjacent piezoelectric AlScN regions for acoustic functionality. Despite being an initial demonstration, these aluminum CPW superconducting resonators exhibit competitive loss performance in both the many-photon and single-photon regimes, achieving quality factors as high as $8.0 \times 10^4$ in the many-photon regime and $7.5 \times 10^4$ in the single-photon regime at 4.175 GHz, the latter corresponding to a coherent lifetime of \(T_1 = 2.9~\mu\mathrm{s}\). Although this performance remains below the state of the art achieved in mature silicon-based superconducting resonator platforms, where process and interface optimization over the past decade have enabled single-photon quality factors exceeding $10^6$, it provides a strong starting point for AlScN-on-SiC and suggests substantial room for improvement. We then take advantage of the remaining high-Sc-content AlScN for electromechanical transduction and phononic guiding. To our knowledge, this work represents the first systematic experimental study of microwave surface phonon propagation loss at milli-Kelvin temperatures in AlScN with such high Sc content. Using SAW delay lines with different propagation lengths, we find that both the Sezawa and Rayleigh mode families exhibit substantially reduced propagation loss at cryogenic temperatures. The extracted losses are $1.02\pm0.49~\mathrm{dB/cm}$ for the Sezawa mode and $3.16\pm0.18~\mathrm{dB/cm}$ for the Rayleigh mode, corresponding to phonon lifetimes of $7.6~\mu\mathrm{s}$ and $3.4~\mu\mathrm{s}$, respectively. Finally, we discuss the dominant mechanisms contributing to phononic loss, including thermoelastic loss, interface roughness, and acoustic leakage and identify clear pathways for future mitigation through material engineering and fabrication optimization. The measured loss values already demonstrate the promising potential of AlScN-on-SiC for low-loss cryogenic microwave phononics, while targeted reduction of the identified loss channels could further advance its performance toward that of state-of-the-art quantum acoustic material platforms.

\section{Superconducting resonators on SiC}

A monolithic quantum acoustic system requires that the substrate support superconducting qubits and microwave circuits with long coherence times. To evaluate the potential of the AlScN-on-SiC platform in this regard, we first fabricate superconducting resonators directly on regions of exposed SiC where the AlScN was selectively removed. During the AlScN etching process, square patches of AlScN were intentionally retained to serve as designated regions for future integration of phononic components. This approach assesses the microwave performance of superconducting qubits on SiC while preserving the piezoelectric layer required for phonon generation and transduction within a monolithic device architecture.

\begin{figure}[htbp]
\includegraphics[width=16cm]{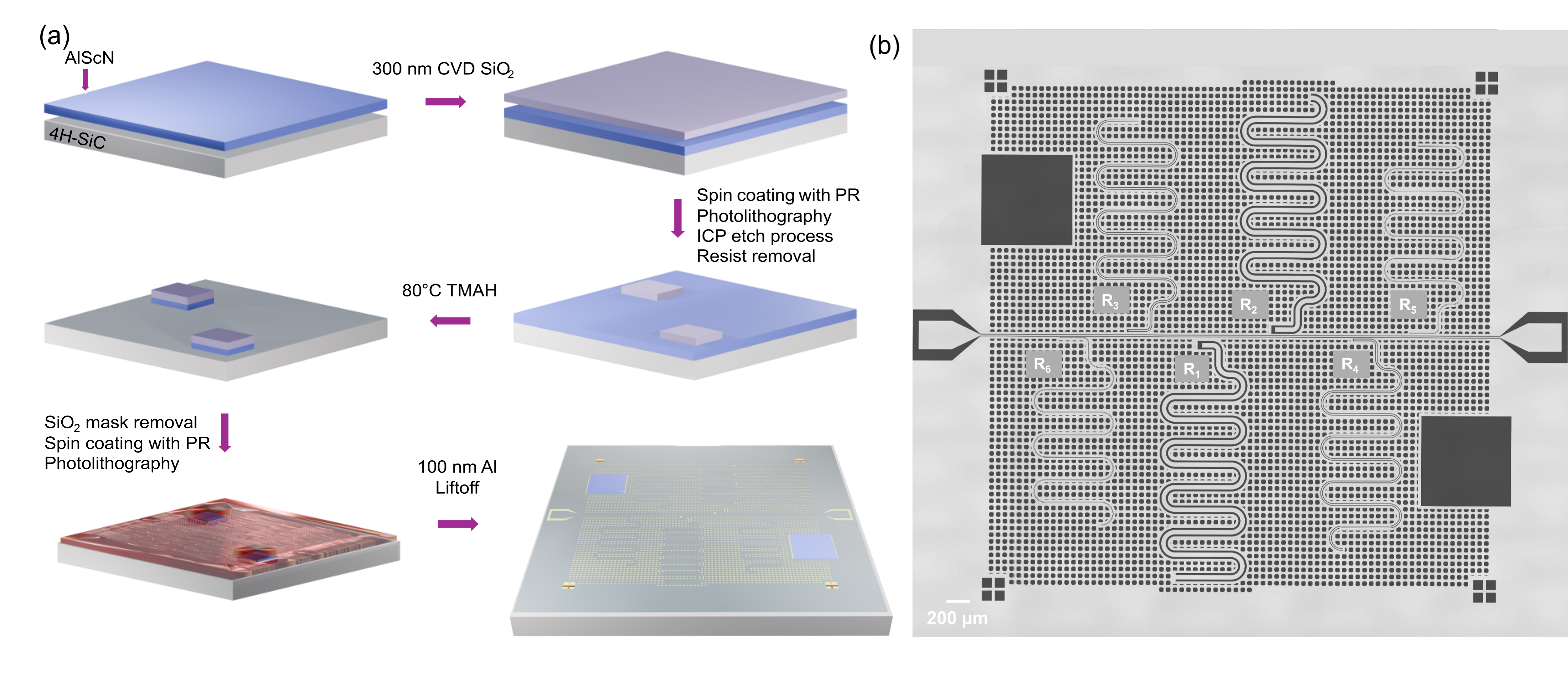}
\renewcommand{\figurename}{Fig.}
\caption{\label{fig:1} (a) Fabrication process flow of the combined AlScN waveguides and superconducting resonators on SiC. (b) Optical image of six superconducting coplanar waveguides (CPWs) with different design parameters as shown in Table I.}
\end{figure}

The fabrication process of the AlScN patches and superconducting resonators is illustrated in Fig.~\ref{fig:1}(a). The process begins with sputter deposition of a $1~\mu\mathrm{m}$ AlScN film with $42\%$ Sc concentration on 4H-SiC, using the same recipe as in our previous studies\cite{Hackett2024S-bandArchitecture,Du2024NearSiC}. A $300~\mathrm{nm}$ silicon dioxide (SiO$_2$) layer is then deposited by PECVD and used as a hard mask for AlScN patterning. A photoresist mask is subsequently patterned onto the SiO$_2$ layer through optical lithography and the exposed SiO$_2$ is etched via an inductively coupled plasma (ICP) process using a CF$_4$/O$_2$/Ar plasma. After removal of the photoresist, the AlScN film is through-etched in a $25\%$ tetramethylammonium hydroxide (TMAH) solution at 80$^{\circ}$C for 25 minutes. Finally, the SiO$_2$ hard mask is removed by immersion in a $5\%$ hydrofluoric acid (HF) solution for 3 minutes. Following removal of the hard mask, fabrication of the superconducting resonators begins. The chip undergoes solvent cleaning prior to photolithographic patterning of the resonators. A layer of AZ5214 photoresist is spin-coated and patterned using optical lithography, followed by development in MIF300 for 30 seconds. After the pattern is defined, a light descum step is then performed to remove any residual photoresist. A 100 nm aluminum film is deposited by electron-beam evaporation, and the resonators are defined via standard lift-off in an acetone bath.

The superconducting resonator chips incorporate six hanger-style $\lambda/4$ coplanar waveguide (CPW) transmission line resonators capacitively coupled to a central feed line, as shown in Fig.~\ref{fig:1}(b). These resonators are designed to operate at frequencies ranging from 4 GHz to 5.5 GHz. To probe whether microwave losses are dominated by surface or bulk SiC contributions, we have selected three different geometries that vary the electric field depth within the SiC substrate. All resonators are designed to have characteristic impedance $Z_c \approx 50 \Omega$. The geometries, from narrowest to widest CPW pattern, are: (1) center conductor c = 4 $\mu m$, gap g = 2 $\mu m$, (2) c = 11 $\mu m$,  g = 5 $\mu m$, and (3) c = 40 $\mu m$,  g = 17 $\mu m$ as summarized in Table I. Each geometry was implemented with two coupling strengths: a weaker coupling yielding $Q_c \approx 10^6$ and a stronger coupling yielding $Q_c \approx 10^5$, providing sensitivity across a broad range of internal losses.

For characterization, the resonator chip was bonded to custom aluminum microwave enclosures and cooled in a dilution refrigerator to a base temperature below 20 mK. Measurement lines were heavily attenuated and filtered to suppress thermal noise and stray radiation, and the sample package was enclosed within a $\mu$-metal shield to reduce background magnetic fields. The internal quality factor, $Q_i$, is extracted from $S_{21}$ measurements performed with a VNA using the procedure described by Probst et al.
\cite{Probst2015EfficientResonators}. 

\begin{figure}[htbp]
\includegraphics[width=14cm]{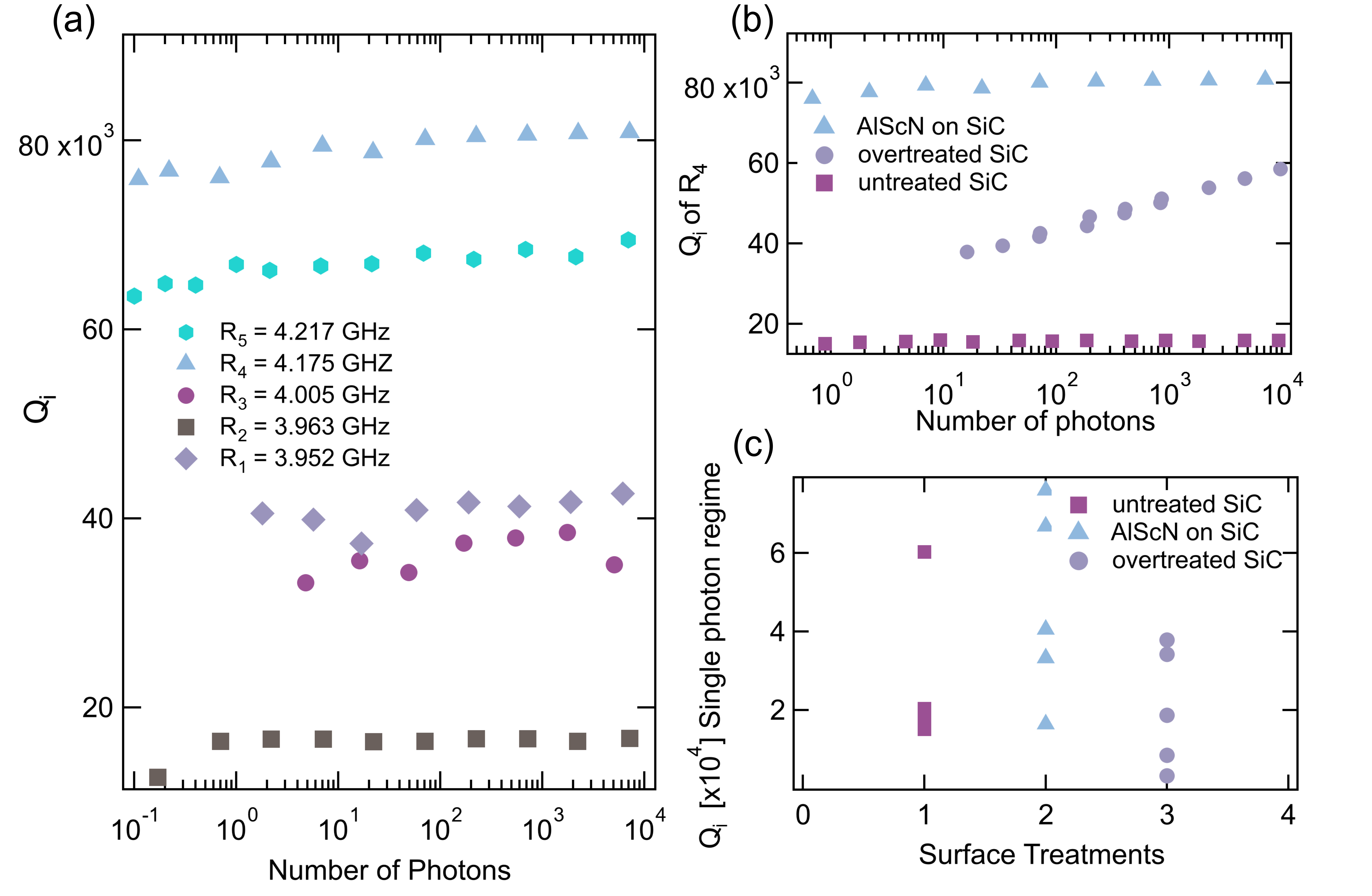}
\renewcommand{\figurename}{Fig.}
\caption{\label{fig:2}(a) Quality factor measurements of five resonators fabricated on AlScN-on-SiC. Among these, the resonator $R_4$ (4.175 GHz) demonstrates the best performance at lower power, featuring a gap of 5 $\mu m$. (b) The quality factor of  $R_4$ improves by a factor of 4 after undergoing the AlScN-on-SiC surface treatment compared to untreated sample. (c) Quality factor measurements at the single-photon limit for the designed resonators subjected to different surface treatment processes.}
\end{figure}

Fig.~\ref{fig:2}(a) presents measurements of $Q_i$ as a function of the average photon number in the resonator $N_{ph}$ for five of the six resonators on our AlScN-on-SiC platform chip, as one resonator did not yield reliable data due to fabrication errors. The medium pattern-width more strongly-coupled resonator ($c = 11 \, \mu \text{m}$, $g = 5 \, \mu \text{m}$ denoted as $R_4$) exhibits the highest internal quality factor with $Q_i \approx 75{,}000$, corresponding to a photon lifetime of $T_1 \approx 2.9~\mu\mathrm{s}$ at $4.175~\mathrm{GHz}$ in the single-photon regime relevant for qubit operations. This geometry likely demonstrates better performance due to a favorable balance between electromagnetic field confinement near the SiC substrate and surface-induced losses. The weakly coupled narrowest-geometry resonator ($c = 4~\mu\mathrm{m}$, $g = 2~\mu\mathrm{m}$, denoted as $R_5$) has $Q_i \approx 65{,}000$, corresponding to a photon lifetime of $T_1 \approx 2.5~\mu\mathrm{s}$ at $4.217~\mathrm{GHz}$. Its narrow geometry confines electric fields closer to the surface of the substrate, increasing surface participation and enhancing sensitivity to surface TLS loss. In contrast, the weakly coupled wide-geometry resonator ($c = 40~\mu\mathrm{m}$, $g = 17~\mu\mathrm{m}$, denoted as $R_1$) exhibits a substantially lower $Q_i \approx 40{,}000$, corresponding to a photon lifetime of $T_1 \approx 1.6~\mu\mathrm{s}$ at $3.952~\mathrm{GHz}$, suggesting significant bulk substrate loss.  The weakly coupled medium-geometry ($c = 11~\mu\mathrm{m}$, $g = 5~\mu\mathrm{m}$) resonator ($R_3$) achieved about the same low-power $Q_i \approx 35{,}000$ ($T_1 \approx 1.4~\mu\mathrm{s}$). The resonator with the lowest $Q_i$ at low power ($R_2$) was the more strongly coupled widest pattern, achieving just $Q_i \approx 12{,}000$ ($T_1 \approx 0.48~\mu\mathrm{s}$).

These trends highlight the critical role of substrate participation and material properties in determining resonator performance. In CPW superconducting resonators, dominant microwave loss channels include surface two-level systems (TLSs) at material interfaces, bulk dielectric loss in the substrate, dissipation from trapped magnetic vortices, and non-equilibrium quasi-particles. In the present measurements, $\mu$-metal shielding suppresses stray magnetic fields well below the vortex-trapping threshold for the CPW geometries employed, while extensive cryogenic attenuation and filtering minimize quasi-particle generation and stray photon loading. Under these conditions, the pronounced degradation of $Q_i$ observed for the widest geometries points to substrate-related loss as the dominant limitation, whereas the narrowest geometries are more consistent with surface TLS-limited behavior.

Taken together, these results indicate that bulk SiC contributes significantly to the total microwave loss in the resonators, while the surfaces of both the SiC substrate and the aluminum films exhibit comparatively lower losses. In addition to conventional dielectric loss, this substrate-related loss may partly arise from the weak but finite piezoelectricity of SiC\cite{Yu2017AcousticCarbide}, which enables microwave phonon energy conversion and introduces an intrinsic radiation loss channel absent in non-piezoelectric substrates such as single-crystal silicon\cite{Zhou2025ObservationSilicon}. Although this effect is expected to be small, it may contribute in geometries with enhanced substrate field participation that can be mitigated through phononic bandgap engineering or resonator designs that minimize electromechanical coupling. Consistent with this interpretation, the intermediate-geometry resonator ($R_4$) achieves the highest $Q_i$, reflecting a favorable balance between surface and substrate participation. In contrast to silicon-based resonators\cite{Earnest2018SubstrateResonators}, where wider geometries typically yield higher $Q_i$, our widest resonators show lower $Q_i$ than other resonators. This is not unexpected given the semiconductor industry's decades-long investment in ultra-high-purity silicon and defect-control processes, which remain underdeveloped for SiC. The present results highlight both the challenges and opportunities associated with SiC, suggesting that further improvements in material quality and electromechanical engineering could improve the coherence time of quantum circuits in this platform\cite{Earnest2018SubstrateResonators,Megrant2012PlanarMillion}.

To further look into the surface treatment correlated loss channels, we also measured resonator $Q_i$ after different surface treatments to ensure that the selective TMAH etching process for AlScN described previously does not overly damage the SiC surface. The three surface treatments that we tried were $(1)$ the standard AlScN etching process with SiC substrate, $(2)$ untreated SiC with no AlScN film, and $(3)$ SiC substrate submerged in TMAH at 80$^{\circ}$C for the same amount of time as the treatment in $(1)$. The same CPW resonator designs were fabricated with equivalent steps for each surface treatment chip. The quality factor of $R_4$ improves by a factor of 4 after treatment $(1)$ compared to treatments $(2)$ and $(3)$ as shown in Fig.~\ref{fig:2}(b). Fig.~\ref{fig:2}(c) further shows $Q_i$ measurements at the single-photon limit for five resonators subjected to different surface treatment processes where all the resonators showcase the same trend. These results indicate that, in the low-power regime, the internal quality factor $Q_i$ of the CPW resonators on this platform is sensitive to the surface preparation and treatment history of the SiC substrate. Such dependence suggests that further optimization of substrate surface processing could provide a viable pathway toward enhancing the microwave coherence of superconducting circuits integrated on the AlScN-on-SiC platform.

These results illustrate that selectively removing AlScN from the AlScN-on-SiC platform can provide exposed SiC regions suitable for low-loss superconducting resonators. Although this represents a first demonstration of superconducting resonators fabricated on selectively etched AlScN-on-SiC, the measured quality factors are already showing promising potential photon lifetime and suggest that further optimization of the etching process, surface preparation, and post-fabrication treatment \cite{Earnest2018SubstrateResonators} could yield substantial improvements. Importantly, this selective integration strategy preserves adjacent AlScN regions as functional piezoelectric elements, enabling phononic devices to be implemented on the same chip. Together with the low-loss cryogenic acoustic delay lines demonstrated in the following section, these results highlight AlScN-on-SiC as a promising monolithic platform for quantum acoustic systems that require the co-integration of superconducting circuits and coherent phononic components.

\begin{table}[htbp]
    \centering
    \begin{tabular}{ccccccc}
    \arrayrulecolor{myColor}\hline
      \textbf{Resonator} & \textbf{Estimated frequency (GHz)} & \textbf{Measured frequency (GHz)} & \textbf{Center ($\mu$m) }& \textbf{Gap ($\mu$m) }& \textbf{Impedance ($\Omega$)} & \textbf{$Q_c$} \\\midrule
        R1 & 4.178 & 3.95 & 40 & 17 & 50 & 1e6 \\
        R2 & 4.432 & 3.96 & 40 & 17 & 50 & 1e5 \\
        R3 & 4.837 & 4.00 & 11 & 5 & 50 & 1e6 \\
        R4 & 5.105 & 4.175 & 11 & 5 & 50 & 1e5 \\
        R5 & 5.404 & 4.21 & 4 & 2 & 50 & 1e6 \\
        \hline
    \end{tabular}
    \caption{Parameters of 5 measured superconducting coplanar waveguides (CPWs) fabricated on SiC}
    \label{tab:my_label}
\end{table}

\section{Phononic losses in AlScN at cryogenic temperature}
\begin{figure}[htbp]
\includegraphics[width=12cm]{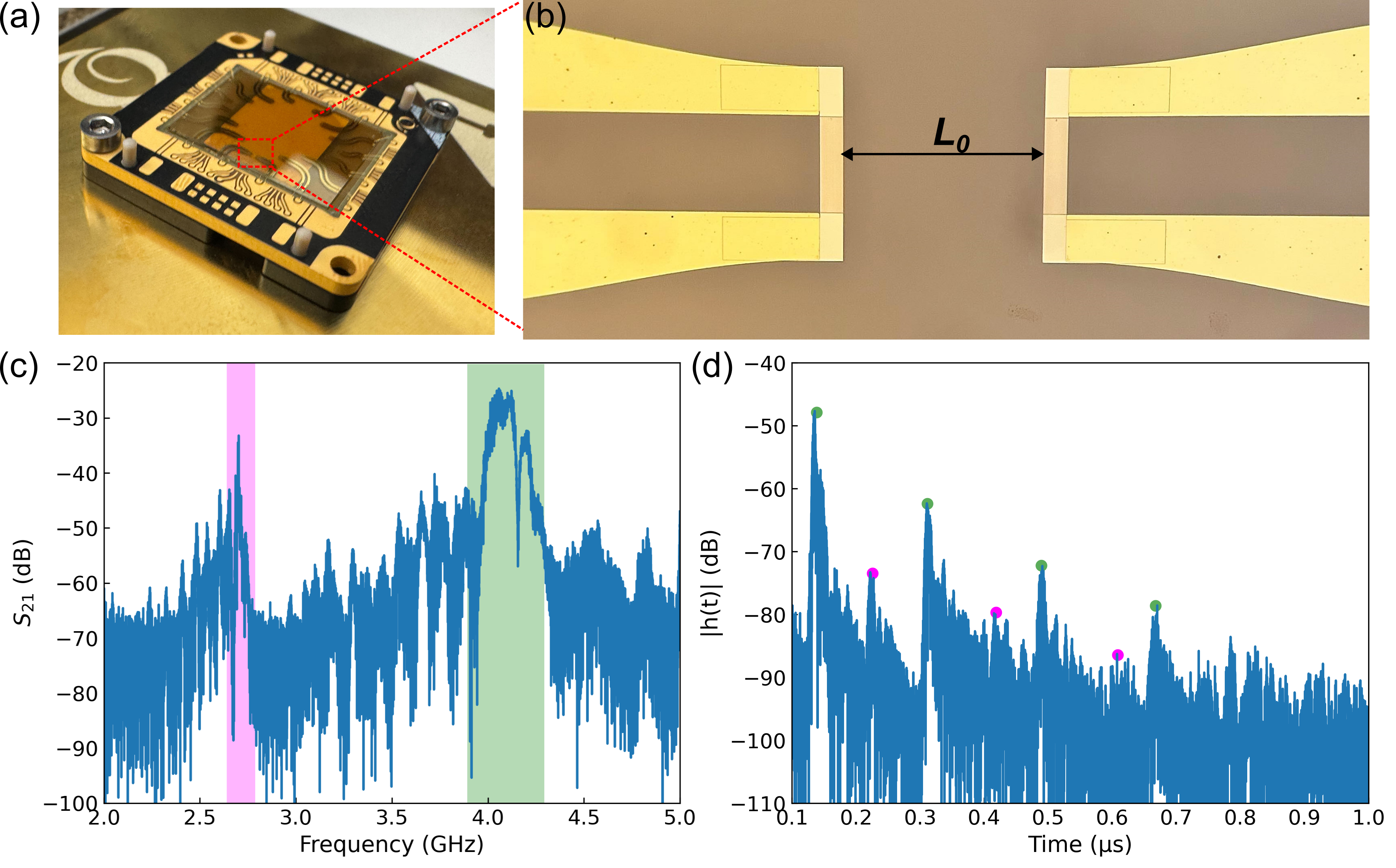}
\renewcommand{\figurename}{Fig.}
\caption{\label{fig:3} (a) A photograph of the AlScN-on-SiC chip with phononic delay lines. (b) A microscope image of the phononic delay line with propagation length $L_0 = 500$ $\mu$m. (c) Example normalized $S_{21}$ transmission data for the phononic delay line with $L_0 = 500$ $\mu$m at 7 mK. The Rayleigh mode and Sezawa mode peaks are marked with pink and green background respectively. (d) The impulse response of the phononic delay line with $L_0 = 500$ $\mu$m at 7 mK, normalized to the full measurement bandwidth. The peaks of the Rayleigh and Sezawa echoes are marked with pink and green dots, respectively.}
\end{figure}

After demonstrating low-loss superconducting resonators on selectively exposed SiC regions, we next evaluate the cryogenic phononic losses of the remaining AlScN piezoelectric layer. We fabricated a series of acoustic delay lines on AlScN-on-SiC with propagation lengths $L_0$ ranging from $500~\mu\mathrm{m}$ to $10000~\mu\mathrm{m}$. The devices were based on a $1~\mu\mathrm{m}$-thick AlScN film deposited on 4H-SiC substrates\cite{Hackett2024S-bandArchitecture,Du2024NearSiC}. The acoustic wavelength was set to $\lambda = 1.6~\mu\mathrm{m}$ to excite a Sezawa mode near $4~\mathrm{GHz}$ with strong electromechanical coupling ($K^2 \approx4.3\%$), matching the frequency range of the superconducting resonators characterized above. As shown in Fig.~\ref{fig:3}(a) and (b), aluminum interdigital transducers (IDTs) with a pitch of 0.8 $\mu$m were patterned with E-beam lithography on a 14 by 11 mm AlScN-on-SiC chip. These IDTs were designed to match the impedance of the RF transmission lines ($50 \Omega$) for efficient generation and detection of the surface phonons. The propagation length $L_0$ of the SAW delay line was defined as the distance between the first fingers of the input and output IDTs. The chip was placed and wire-bonded to a Quantum Machines QBoard-II cryogenic sample holder and then cooled down to 7 mK in a dilution refrigerator. The S-parameters of the delay lines were measured using a Keysight P9373B vector network analyzer (VNA). The input power was set to -80 dBm. To increase the SNR, a Low Noise Factory high-electron-mobility transistor (HEMT) amplifier LNF-LNC0.3\_14B was utilized with 40 dB of gain for the Rayleigh mode and 39.5 dB of gain for the Sezawa mode. Assuming the losses are symmetric on the input and output lines, we achieve approximately -100 dBm of on chip power for the Rayleigh mode and -92.5 dBm for the Sezawa mode. As shown in Fig.~\ref{fig:3}(c), the transmission spectrum exhibits two distinct peaks at 2.69 GHz and 4.05 GHz, corresponding to the Rayleigh and Sezawa SAW modes, respectively. Fig.~\ref{fig:3}(d) displays the time-domain impulse response from the inverse Fourier transform of the $S_{21}$ spectrum, showing a primary transmission peak followed by a series of echoes attributed to reflections off the IDT fingers. 

\begin{figure}[htbp]
\includegraphics[width=12cm]{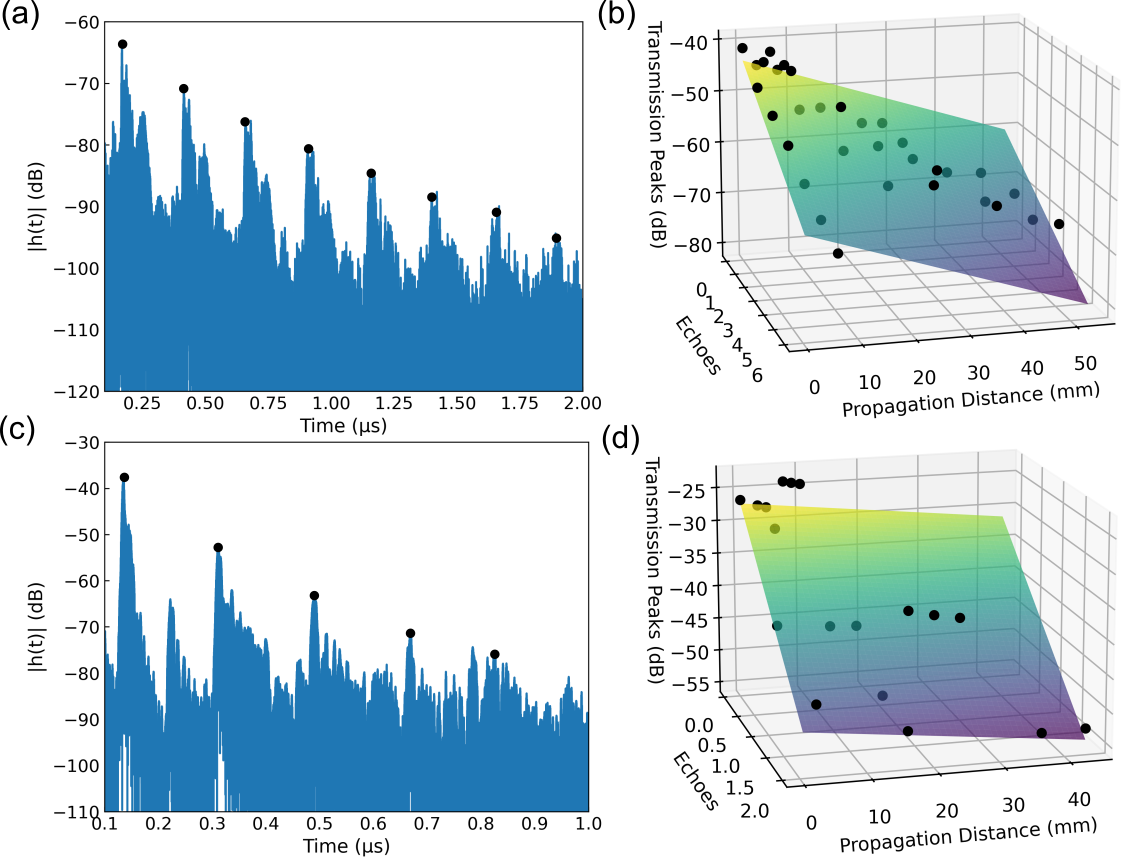}
\renewcommand{\figurename}{Fig.}
\caption{\label{fig:4} (a) The impulse response $|h|$ of the Rayleigh mode in the 500 $\mu$m delay line at 7 mK, normalized to the filter passband. (b) The amplitudes of the Rayleigh mode echo peaks are fitted into a plane against the number of echoes and propagation distances. (c) The impulse response $|h|$ of the Sezawa mode in the 500 $\mu$m delay line at 7 mK, normalized to the filter passband. (d) The amplitudes of the Sezawa mode echo peaks are fitted into a plane against the number of echoes and propagation distances.}
\end{figure}

To isolate the propagation losses of the Rayleigh and Sezawa modes, we employed a time-gating procedure on the measured $S_{21}$ data of the corresponding phononic mode. In the time-domain responses of the Rayleigh and Sezawa modes, shown in Fig.~\ref{fig:4}(a) and (c), respectively, we identified a sequence of echo peaks corresponding to multiple acoustic transits between the IDTs. Each echo peak results from reflection at the transducers, with the power of the $n$-th echo expressed as
\begin{equation}
    P_{\mathrm{lin}} = T^2 R^{2n} e^{-\alpha_{\mathrm{lin}} L},
\end{equation}
where $T$ is the IDT conversion efficiency, $R$ is the acoustic power reflection coefficient, $n$ is the echo number, and $L = (2n+1)L_0$ is the total propagation distance\cite{Mayor2021GigahertzSapphire}. Taking the logarithm, the transmitted acoustic power in dB can be expressed as
\begin{equation}
    P_{\mathrm{dB}} = 20\log_{10}T + 20n\log_{10}R - 10\alpha_{\mathrm{lin}}(\log_{10}e)L,
\end{equation}
where $\alpha_{\mathrm{dB}} = 10\alpha_{\mathrm{lin}}\log_{10}e$ represents the propagation loss in dB per unit length. The dB power levels of successive echoes were Fourier transformed from gated time-domain echo peaks and fitted to planar surfaces in the 3D space spanned by $L$, $n$, and $P_{\mathrm{dB}}$, as shown in Fig.~\ref{fig:4}(b) and (d). From these planar fits, we obtain a propagation loss of $1.02\pm0.49~\mathrm{dB/cm}$ for the Sezawa mode and $3.16\pm0.18~\mathrm{dB/cm}$ for the Rayleigh mode. Compared with previously reported room-temperature results\cite{Deng2025MonolithicDevices}, these cryogenic losses are reduced by approximately $50\times$ for the Sezawa mode and $30\times$ for the Rayleigh mode. These results show that AlScN on SiC can support low-loss SAW propagation at cryogenic temperatures while preserving the strong electromechanical coupling enabled by high Sc concentration AlScN.

From the time-domain impulse response, we also extract the group velocity of the two modes using $v_g = 2L_0/\Delta t$, where $\Delta t$ is the time delay between successive echoes. By calculating the time delays across different delay lines, we obtain a Sezawa-mode group velocity of approximately $v_{g,S} = 5629~\mathrm{m/s}$ and a Rayleigh-mode group velocity of approximately $v_{g,R} = 4004~\mathrm{m/s}$. Using these measured velocities and propagation losses, the theoretical mechanical quality factor of an AlScN phononic cavity on SiC can be estimated as
\begin{equation}
    Q = \frac{\omega}{v_g\alpha_{\mathrm{lin}}},
\end{equation}
yielding $Q\approx190000$ for the Sezawa mode and $Q\approx58000$ for the Rayleigh mode at cryogenic temperatures. These quality factors correspond to phonon lifetimes of approximately $\tau_S \approx 7.6~\mu\mathrm{s}$ for the Sezawa mode and $\tau_R \approx 3.4~\mu\mathrm{s}$ for the Rayleigh mode. These lifetimes are long enough for many quantum applications\cite{Krenner2026TheRoadmap}, including quantum random access memory (RAM)\cite{Hann2019Hardware-EfficientSystems}, bosonic encoding of qubits\cite{Diringer2024Conditional-notQubit}, and quantum entanglement between SAW phonons\cite{Wollack2022QuantumResonators}.

While the actual phonon lifetime in fabricated phononic resonators may be affected by the resonance mode profile, radiative loss into the SiC substrate, and mirror leakage due to fabrication tolerances, the measured delay-line losses provide an important upper-bound estimate of the propagation-loss-limited performance. The observed propagation losses likely arise from a combination of multiple loss channels in the AlScN-on-SiC platform at cryogenic temperatures. Surface roughness in sputtered AlScN films can induce elastic scattering of SAW modes, particularly in the multi-gigahertz regime where the acoustic wavelength becomes sensitive to nanoscale roughness. This mechanism is expected to affect the Rayleigh mode more strongly because of its shallower energy confinement near the surface, whereas the Sezawa mode exhibits reduced sensitivity due to its deeper mode profile. Consistent with this picture, the time-domain responses reveal significantly stronger acoustic reflection for the Rayleigh mode, with $R_{\mathrm{Rayleigh}} = 0.47$, than for the Sezawa mode, with $R_{\mathrm{Sezawa}} = 0.071$, possibly due to the Rayleigh mode's enhanced susceptibility to surface perturbations and mass loading from the IDT fingers. Intrinsic thermoelastic loss mechanisms, such as Akhiezer damping\cite{Kunal2011AkhiezerNanostructures}, may also contribute to the acoustic attenuation particularly at elevated temperatures. The strong reduction in propagation loss observed at cryogenic temperatures compared with room-temperature measurements may therefore be partially associated with the suppression of Akhiezer damping\cite{Lin2026ExperimentalNiobate}. However, because temperature dependent measurements were not performed in this work, the relative contribution of this mechanism cannot be quantitatively separated from other loss channels. Additional loss may arise from strain-coupled two-level systems at cryogenic temperatures, similar to other piezoelectric heterostructure systems\cite{Wollack2021LossTemperature,Gruenke-Freudenstein2025SurfaceResonators}. These TLSs may be associated with disordered regions, native oxides, amorphous interfacial layers, or grain boundaries. In our film stack, an AlN seeding layer is used to promote high-quality AlScN growth on SiC. While this layer improves film texture of directly sputtered high-Sc-content AlScN on SiC, it also introduces additional interfaces between the material layers. These buried interfaces could host interfacial TLS and may also provide additional scattering or leakage channels for surface phonons.

To further improve AlScN-on-SiC as a quantum acoustic platform, the cryogenic phononic loss can be reduced by optimization of AlScN growth conditions, reduction of surface roughness, improved control of interfacial oxide layers, surface passivation, and engineering of the Sc concentration to balance electromechanical coupling and crystalline quality. Phononic cavity designs with optimized acoustic mirrors, reduced substrate radiation, and improved mode confinement should further increase the achievable mechanical quality factors. Although these optimization directions require further experimental validation, the present cryogenic delay-line measurements provide an important benchmark for cryogenic phononic loss in AlScN-on-SiC and support its potential use in low-loss quantum phononic devices integrated with superconducting circuits.

\section{Outlook for Quantum Acoustic Architectures}

The demonstrated superconducting and phonon lifetimes already place this platform within reach of being able to achieve high performance for several different quantum acoustic architectures. A superconducting microwave lifetime of approximately $3~\mu\mathrm{s}$ and a piezoelectric phonon lifetime of approximately $8~\mu\mathrm{s}$ only require coupling rates of order 100 kHz to achieve strong coupling. However, the electromechanical coupling coefficient of the Sezawa mode used here has a theoretical upper bound of 8\% (4.3\% demonstrated), which is on par with suspended LiNbO$_3$ quantum acoustic cavities demonstrated previously\cite{Chiappina2023DesignTransducer},  which means this platform should in principle enable coherent coupling rates of order 10 MHz, several orders of magnitude higher than the threshold rate.  This would immediately enable coherent photon-phonon exchange and detection, sufficient for proof-of-principle demonstrations of phonon-mediated quantum state transfer\cite{Bienfait2019Phonon-mediatedEntanglement}, quantum communication with acoustic delay-lines\cite{Dumur2021QuantumPhonons}, and phononic interferometers or beam-splitter networks coupled to superconducting quantum circuits. 

\begin{figure}[htbp!]
    \centering
    \renewcommand{\figurename}{Fig.}
    \includegraphics[width=0.75\linewidth]{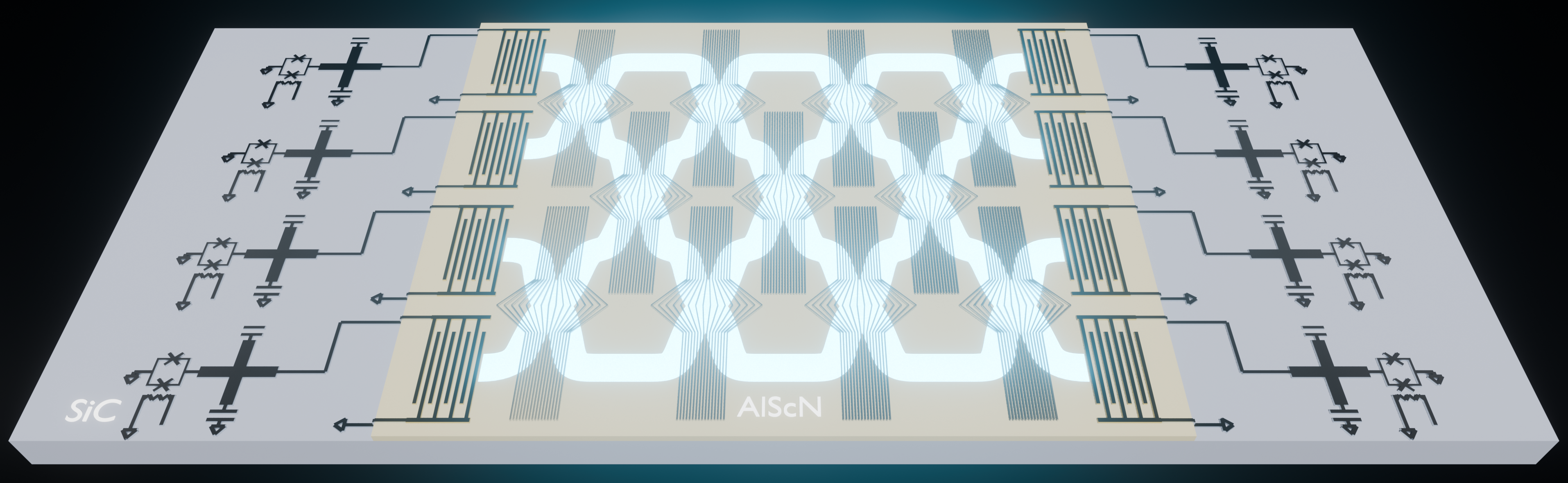}
    \caption{Conceptual picture of a phononic static Gaussian boson sampling circuit in AlScN-on-SiC.}
    \label{fig:5}
\end{figure}

Further improvement in both photon and phonon lifetimes would enable more complex quantum acoustic architectures, including multiple sequential qubit-phonon exchange operations, lower-loss propagation through phononic networks, and higher-fidelity interference between itinerant phonons. These capabilities are essential for larger phononic quantum circuits, such as static Gaussian boson sampling networks shown in Fig.~\ref{fig:5}, where nonclassical phononic states generated by superconducting devices on SiC propagate through fixed acoustic beam splitters and delay lines on AlScN before being measured by superconducting detectors. In the longer term, such low-loss integrated phononic circuits could form a hardware layer for MBQC. Continued reduction of dielectric, interfacial, and acoustic propagation losses is therefore a direct path toward scalable hybrid quantum acoustic processors.

\section{Conclusion}
In conclusion, we investigate AlScN thin films with $42\%$ Sc concentration on 4H-SiC as a monolithic platform for quantum acoustic systems that integrate superconducting circuits with piezoelectric phononic devices. We first establish a selective AlScN removal process that exposes low-loss SiC regions while preserving adjacent AlScN areas for acoustic transduction. Superconducting CPW resonators fabricated on the exposed SiC exhibit internal quality factors of $7.5\times10^4$ in the single-photon regime. Although the estimated photon lifetime of $T_1 \approx 2.9~\mu\mathrm{s}$ remains below that of superconducting resonators on more mature platforms such as Si, this first demonstration proves that the selective fabrication process of the AlScN-on-SiC is compatible with superconducting quantum circuits. We then characterize SAW delay lines fabricated on the $42\%$ Sc concentration AlScN and observe low cryogenic propagation losses for both Rayleigh and Sezawa modes at microwave frequencies. These results show that the same AlScN-on-SiC chip can simultaneously support high-Q superconducting resonators and long-lived surface phonons.

The present work represents an initial demonstration of this material platform for quantum acoustic applications rather than an optimized device implementation. The measured acoustic and microwave losses may still be affected by surface roughness, interfacial disorder, TLS loss, substrate radiation, and crystal defects associated with high Sc AlScN. Continued improvements in film growth, selective etching, surface treatment, interface control, and phononic confinement are therefore expected to further enhance both superconducting and acoustic coherence time. Nevertheless, the coexistence of low-loss superconducting resonators and low-loss piezoelectric SAW propagation on a selectively patterned AlScN-on-SiC heterostructure establishes an important starting point for monolithically integrated cQAD and MBQC architectures in which superconducting qubits, microwave resonators and phononic devices can be simultaneously designed for quantum information processing systems and quantum networks.

\begin{acknowledgments}

This article is based on research sponsored in part by the Defense Advanced Research Projects Agency (DARPA) through a Young Faculty Award (YFA) under grant D23AP00174-00. The views and conclusions contained herein are those of the authors and should not be interpreted as necessarily representing the official policies or endorsements, either expressed or implied, by DARPA, the Department of the Interior, or the US Government. This material is based upon work supported by the U.S. Department of Energy, Office of Science, National Quantum Information Science Research Centers, Quantum Systems Accelerator (Award No. DE-SCL0000121). This work was performed, in part, at the Center for Integrated Nanotechnologies, an Office of Science User Facility operated for the U.S. Department of Energy, Office of Science. This work was also carried out, in part, at Sandia National Laboratories, a multimission laboratory managed and operated by National Technology and Engineering Solutions of Sandia, LLC, a wholly owned subsidiary of Honeywell International, Inc., for the U.S. Department of Energy’s National Nuclear Security Administration under Contract No. DE-NA-003525. This work was also carried out in part at the Singh Center for Nanotechnology, which is supported by the NSF National Nanotechnology Coordinated Infrastructure Program under grant NNCI-2025608. This paper describes objective technical results and analysis. Any subjective views or opinions that might be expressed in the paper do not necessarily represent the views of the U.S. Department of Energy or the U.S. Government.
\end{acknowledgments}


\bibliography{apl_cqad.bib}

\end{document}